\documentclass[10pt,journal]{IEEEtranTCOM}

\usepackage[english]{babel}
\usepackage[usenames]{color}
\usepackage[cp1250]{inputenc}
\usepackage{amsfonts}
\usepackage{amsthm}
\usepackage{graphicx}
\usepackage{epsfig}
\usepackage{mathrsfs}
\usepackage{amsmath}
\usepackage{algorithm}
\usepackage{algorithmic}
\usepackage{hyperref}

\theoremstyle{plain}

\begin{document}
\newcommand{\bea}{\begin{eqnarray}}
\newcommand{\eea}{\end{eqnarray}}
\newcommand{\be}{\begin{equation}}
\newcommand{\ee}{\end{equation}}
\newcommand{\beas}{\begin{eqnarray*}}
\newcommand{\eeas}{\end{eqnarray*}}
\newcommand{\bs}{\backslash}
\newcommand{\bc}{\begin{center}}
\newcommand{\ec}{\end{center}}
\def\SC {\mathscr{C}}

\title{Context binning, model clustering and adaptivity\\ for data compression of genetic data}
\author{\IEEEauthorblockN{Jarek Duda}\\
\IEEEauthorblockA{Jagiellonian University,
Golebia 24, 31-007 Krakow, Poland,
Email: \emph{dudajar@gmail.com}}}
\maketitle

\begin{abstract}
Rapid growth of genetic databases means huge savings from improvements in their data compression, what requires better inexpensive statistical models. This article proposes automatized optimizations e.g. of Markov-like models, especially context binning and model clustering. While it is popular to just remove low bits of the context, proposed context binning automatically optimizes such reduction as tabled:  \texttt{state=bin[context]} determining probability distribution, this way extracting nearly all useful information also from very large contexts, into a relatively small number of states. The second proposed approach: model clustering uses k-means clustering in space of general statistical models, allowing to optimize a few models (as cluster centroids) to be chosen e.g. separately for each read. There are also briefly discussed some adaptivity techniques to include data non-stationarity.
\end{abstract}
\textbf{Keywords}: data compression, genetic data, (hidden) Markov model, k-means clustering, adaptivity, non-stationarity
\section{Introduction}
We live in times of rapid growth of bioinformatics, e.g. for individual treatment of each patient based on sequencing in the precision medicine approach~\cite{prec}. However, sequencing data for a single person can reach terabytes, requiring huge databases - which optimizations by improved data compression techniques can lead to significant savings.

\begin{figure}[t!]
    \centering
        \includegraphics{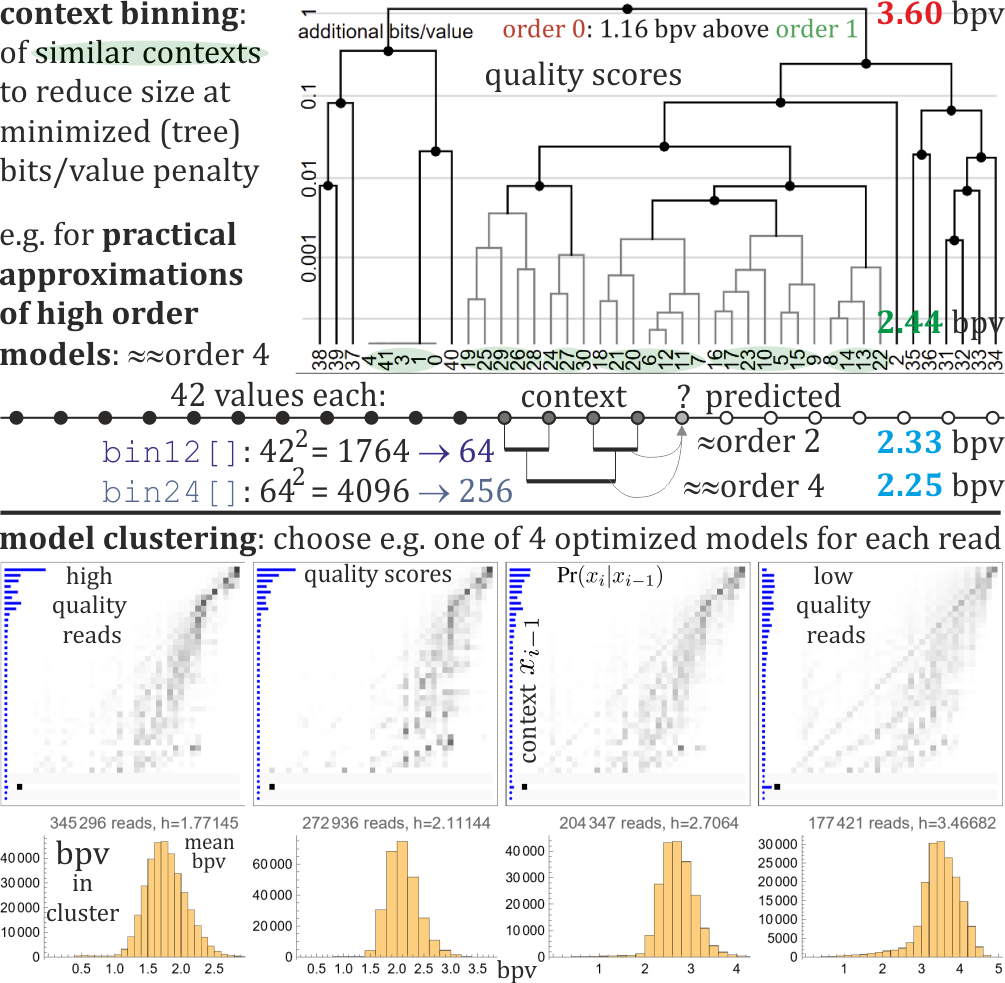}
        \caption{Two main approaches proposed in this article. \textbf{Top}: context binning automatically merging subsets of contexts into a smaller number of states \texttt{context}$\to$ \texttt{state=bin[context]} determining probability distribution for the currently processed symbol. It is optimized to minimize bits/value distance from complete model distinguishing all contexts, and can allow e.g. for inexpensive approximations of high order models. \textbf{Bottom}: model clustering - different e.g. reads can have slightly different statistics, suggesting to allow to choose one of e.g. 4 shown order 1 models (as cluster centroids), optimized with k-means clustering applied to space of  models. We can e.g. see separate treatment of low and high quality reads.}
       \label{intr}
\end{figure}

Dependencies in this type of data are relatively weak. Beside Burrows-Wheeler transform~\cite{BW}, there are usually dominating simple techniques like entropy coding, RLE (run-length encoding: representing blocks of identical values), extending toward Markov models~\cite{cram} - optimizations of which we are focused here, briefly presented in Fig. \ref{intr}.

One optimization direction is toward higher order models and exploiting additional contexts like position, where the basic difficulty is exponential size growth with order. Proposed context binning allows for its practical approximations by automatically merging similar contexts, extracting nearly all useful information also from very large contexts, e.g. from $2^l$ previous values in $l$ table uses per symbol.

Second parallel optimization direction is varying models accordingly to local statistics. Briefly mentioned standard way is adaptation to include non-stationarity. Genetic data often has characteristic situation: of large number of reads treated as independent, what suggests using various models for different reads. While reads are too short for storing entire models in header, we can use clustering philosophy - finding a few optimized models (cluster centroids), and choosing one of them e.g. with a few bits written in header of read.

The proposed approaches are inexpensive from decoder and encoder side to directly apply. However, their optimization for actual data might be more costly - for example it could be done once e.g. for each sequencer model (like Illumina HiSeq 2000), and then used as default - e.g. 20 models for sequencer, compressor chooses 4 best of them for given file, then 2 bits per header choose one of 4 models.

Diagrams in this article contain analysis based on publicly available data - mainly first 1 million reads (all of 101 length) from FASTQ file ERR174310 (Whole Genome Sequencing of human, Illumina HiSeq 2000) downloaded from NCBI database\footnote{https://www.ncbi.nlm.nih.gov/sra/ERR174310}. Additionally, for long subsequent base sequence, nonstationarity analysis was made for human Chromosome 1 using FASTA file from ensembl.org webpage\footnote{http://ftp.ensembl.org/pub/release-105/fasta/homo\_sapiens/dna/}.

\section{Context binning}
We would like to compress $\{x_i\}_{i=1..N}$ sequence from size $|\mathcal{A}|=m$ alphabet ($x_i\in\mathcal{A}$), e.g. $m=4$ for ATCG nucleotides, $m\sim 40$ quality scores, or some grouped symbols e.g. $m\sim 4\cdot 40 =160$ for (nucleotide, its quality score) grouped to include their statistical dependencies (\texttt{(qscore<<2)|base}).

Popular natural approach is encoding symbol-by-symbol, predicting conditional distribution of the next symbol based on the previous ones: $\textrm{Pr}(x_i|x_{i-1}\ldots x_1))$ and maybe some additional contexts like position. Assuming accurate entropy coding like Arithmetic Coding (AC)~\cite{AC} or Asymmetric Numeral Systems (ANS)~\cite{ANS}, symbol of probability $p$ needs asymptotically $\lg(1/p)$ bits ($\lg\equiv \log_2$). While in prefix codes like Huffman we approximate this $\lg(1/p)$ with a natural numbers of bits ($p$ with natural powers of 1/2), AC/ANS asymptotically include also fractional bits.

Hence we can focus on conditional probability modelling with final evaluation as compression ratio given by:
\be \frac{1}{N} \sum_{i=1}^N \lg(1/\textrm{Pr}(x_i|x_{i-1}\ldots x_1))\quad \textrm{bpv (bits/value)} \ee

In order $l$ Markov model we restrict history to $l$ previous values (with some special treatment of first $l$ values): $\textrm{Pr}(x_i|x_{i-1}\ldots x_{i-l})$ for $x_i\in\mathcal{A}$, $i>l$. The main difficulty is exponential growth of size of such models: requiring $m^l$ of size $m$ probability distributions. We can also add to context more looking relevant information, e.g. in FQZComp~\cite{cram} there is also encoded position in read, number of changes of quality score.

Let us discuss here automatic merging of multiple such contexts into a smaller number of states providing nearly the same compression ratio at much smaller computational cost.

\subsection{Direct context binning}
Let $C$ be a set of contexts, e.g. $C=\mathcal{A}^l$ for order $l$ model, each defining a probability distribution $P_c$ for $c\in C$, $i>l$:
\be \textrm{Pr}(x_i|x_{i-1}\ldots x_{i-l})=P_c(x_i)\quad \textrm{for}\quad c=(x_{i-1},\ldots, x_{i-l}) \ee
Taking all $l+1$ length windows: $$W=(x_i,x_{i-1},\ldots,x_{i-l})_{i=l+1..N}$$ $$\textrm{denote}\quad [0,1] \ni p_c =|\{w\in W: w_{2..l+1}=c\}|/|W|$$ as probability of context $c$ $\left(\sum_{c\in C} p_c =1\right)$,
$$P_c(X)=\frac{1}{|W|\,p_c}\{w\in W: w_1=x, w_{2..l+1}=c\}|$$ as probability distribution for new symbol after this context $\left(\sum_{x\in \mathcal{A}} P_c(x) =1\right)$. Empty contexts $p_c=0$ can be handled e.g. by adding some tiny $\epsilon>0$ to joint distribution on $W$.

Now the optimized rate in bits/symbol can be written as
\be R=\sum_{c\in C} p_c H(P_c)\qquad \textrm{bpv (bits/value)}\ee
$$\textrm{for}\quad H(P)=-\sum_{x\in \mathcal{A}} P(x)\lg\left(P(x)\right) $$
being Shannon entropy after context $c$ of probability $p_c$.\\

\begin{figure}[t!]
    \centering
        \includegraphics{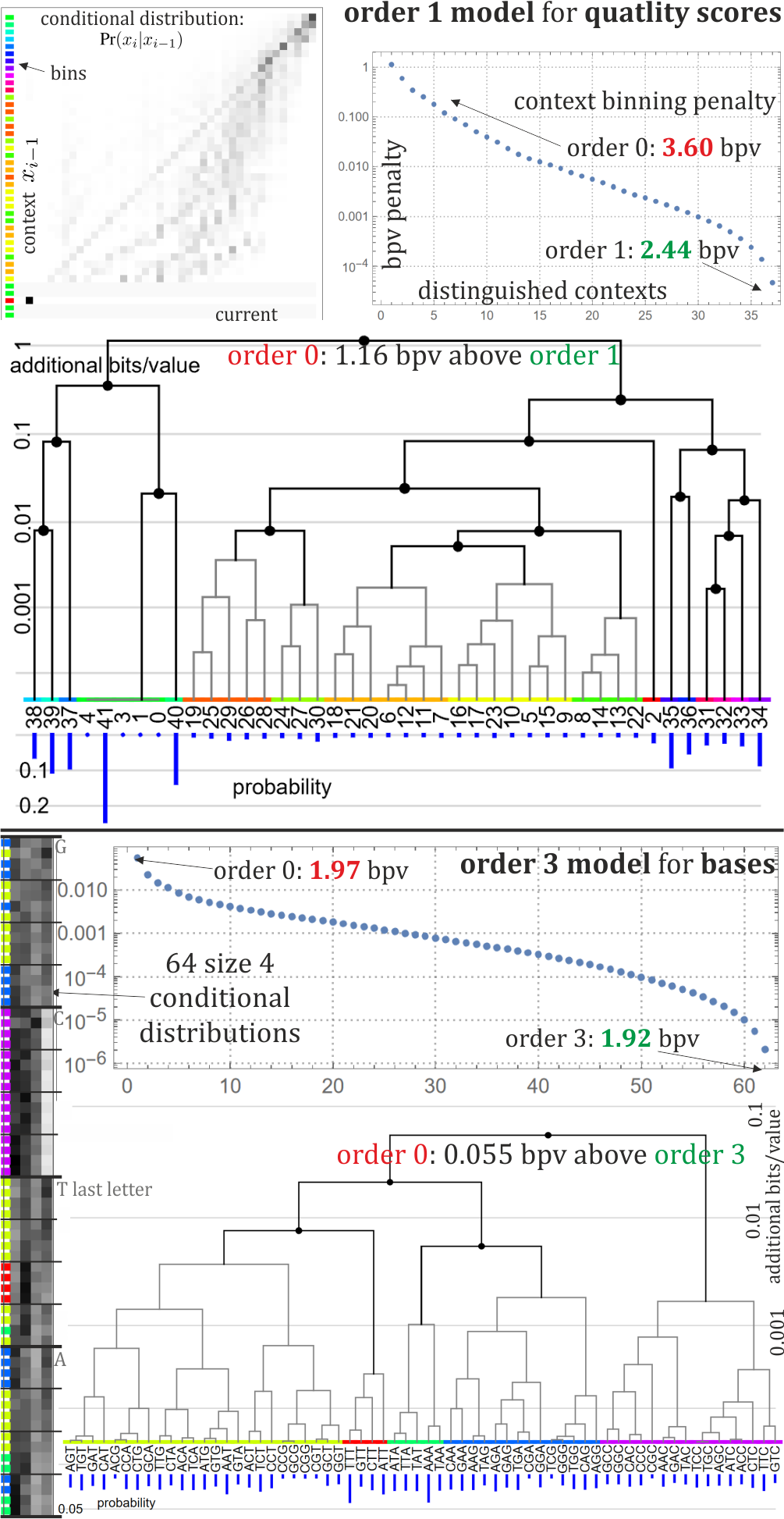}
        \caption{Context binning for order 1 model for quality scores (top) and order 3 for bases (bottom). Distinguishing all previous 42 values (64 for bases) we need 2.44 bpv (1.97 for bases) - using presented conditional distributions. Completely neglecting this context we need 1.16 bpv more: 3.60 bpv (0.055, 1.92 for bases). The top plots and trees show optimized intermediate options: starting with singletons we merge subsets of contexts leading to the smallest bpv penalty constructing the tree, then merge leafs of its subtrees into smaller contexts. Presented context binning into 17 (5 for bases, marked with colors) states was chosen not to exceed 0.01 bpv from complete context models.}
       \label{binning}
\end{figure}

As context binning, we would like to merge similar contexts - replace $C$ set of contexts with $\bar{C}$ set of disjoint subsets of $C=\bigsqcup_{s\in \bar{C}}\, s$.

Having two disjoint subsets of contexts $s,r\subset C$, $s\cap r=\emptyset$,  probability of their union and bits/value cost of their merging are:
$$p_{s\cup r}=p_s+p_r\qquad \qquad P_{s\cup r} =\frac{p_s P_s +p_r P_r}{p_s + p_r}$$
\be \Delta_{s,r} = (p_s+p_r)H\left(P_{s\cup r}\right)-p_s H(P_s) - p_{r}H(P_r)\label{dif}\ee

For optimization of such division, we can start with subsets as singletons $s=\{c\}$ for $c\in C$, with $p_s=p_c$, $P_s=P_c$. Then try to merge them in a way minimizing cost (\ref{dif}) as increase of bits/symbol rate for not recognizing these two subsets of contexts - e.g. in greedy way as in used here Algorithm \ref{bin}.

\begin{figure}[t!]
    \centering
        \includegraphics{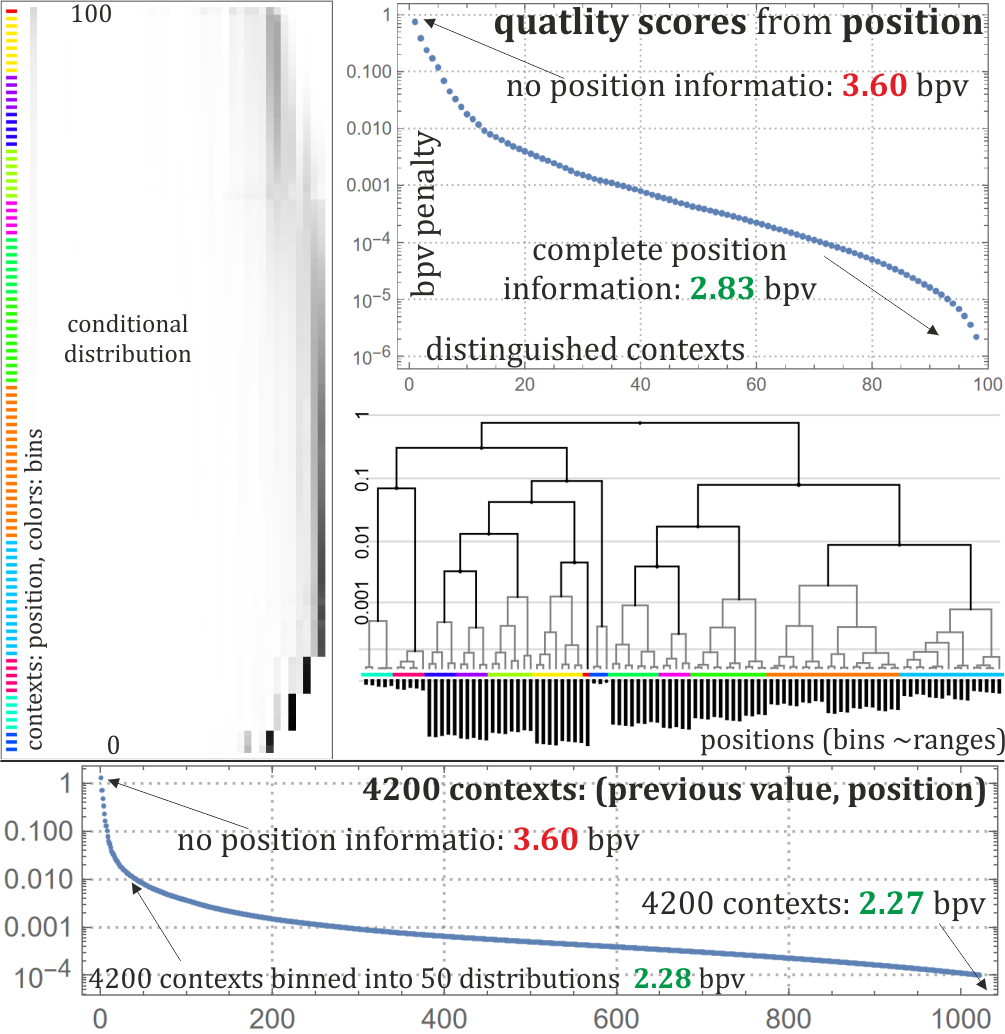}
        \caption{Using position as additional information (used by FQZComp~\cite{cram}). In our case (ERR174310) it is easy to directly include as all reads have the same length (101), and we can see that indeed probability distributions strongly vary with positions here (top left). There is shown binning not to exceed 0.01 bpv penalty: in tree and colors on the left - as we could expected, the found bins turned out being ranges of positions. Bottom: combining with Markov model, using also previous value as context. We could also further nest with other contexts e.g. values in earlier positions.}
       \label{pos}
\end{figure}

\begin{algorithm}[htbp]
\footnotesize{
\caption{Greedy search for context binning}
\label{bin}
\begin{algorithmic}
\STATE \texttt{nodes} $=\{\{c\}:c\in C\}$  \qquad\qquad\qquad\COMMENT{start with leafs as singletons}
\STATE \texttt{available}=\{True,\ldots,True\} of length $|C|$ \qquad\qquad\COMMENT{not used yet}
\STATE initialize HEAP: insert with PUSH, POP retrieves lowest cost new node
\FOR{$c\neq c'\in C$}
\STATE PUSH$((\Delta,\{c\},\{c'\}))$ using ($\ref{dif}$) cost of $\{c\},\{c'\}\to \{c,c'\}$ merge
\ENDFOR
\FOR{$i=1$ to $|C|-1$}
\STATE \{\texttt{cur},\texttt{left},\texttt{right}\}=POP \qquad\COMMENT{retrieve lowest cost node candidate}
\WHILE{\texttt{left} or \texttt{right} not \texttt{available}}
\STATE \{\texttt{cur},\texttt{left},\texttt{right}\}=POP \qquad\quad\COMMENT{skip already merged nodes}
\ENDWHILE
\STATE mark \texttt{left} and \texttt{right} as not \texttt{available}
\STATE append True to \texttt{available}
\STATE append \texttt{cur} to \texttt{nodes}, link \texttt{left}, \texttt{right} as its children in tree
\FOR{all \texttt{available} \texttt{node}}
\STATE PUSH(($\Delta$,\texttt{cur},\texttt{node})) using (\ref{dif}) to calculate merging cost $\Delta$
\ENDFOR
\ENDFOR
\end{algorithmic}
}
\end{algorithm}
This way we get tree of subsets of $C$ as in example in Fig. \ref{binning}, \ref{pos} - w can choose the actual context binning e.g. by starting with the root, and extending with the lowest cost neighboring nodes until reaching some threshold e.g. the number of binned contexts, total bpv distance from using entire context, bpv cost of adding new node, etc.

As there is freedom of bin enumeration, we can e.g. optimize it to be able to get denser and sparser binning from single table use: as \texttt{bin[]} and e.g. \texttt{bin[]>>3}, this way merging 8 smaller bins into larger ones. To optimize it, we can start with finding larger number of bins. Then starting with these bins as contexts, optimize for smaller number of bins - finally grouping e.g. 8 small into 1 large.

\subsection{Nested context binning}
Above direct context binning allows e.g. to approximately represent 2 or a few previous values with a more reasonable number of states, also combine with other contexts like positions in Fig. \ref{pos}.  We can e.g. modify fast order 1 rANS implementation\footnote{https://github.com/jkbonfield/rans\_static} by adding single table use to get more valuable e.g. 1 byte context. E.g. for order 2 as \texttt{state=bin12[(pval<<6)|val]} for \texttt{pval}$=x_{i-2}$, \texttt{val}$=x_{i-1}$.

We  can also combine multiple binnings e.g. to get approximated higher order models. Like in Fig. \ref{hier} for order 4 model: using $\textrm{Pr}(x_i|x_{i-1},x_{i-2},x_{i-3},x_{i-4})$ we can first perform binning of pairs of values using \texttt{bin12[]}, then use second binning to group representations of both pair using \texttt{bin24[]}. This way we use two approximations reducing the original context size $42^4=3\,111\,696$ into more reasonable e.g. 256 number of states (convenient 1 byte) determining probability distribution for $x_i$.

This is the simplest, least expensive way (symmetric) to combine context binnings. In practice contributions of further contexts is usually weakening - suggesting to use smaller numbers of bins for them (asymmetric) or even depending on nearer bins (hierarchical), e.g. to allow for larger orders:
\begin{enumerate}
\item Symmetric: use identical context binning for all windows of given size (as discussed) e.g. 2 and 4 in Fig. \ref{hier}, what is the least expensive (reusing previous binnings): e.g. $2^l$ order for $l$ table uses per symbol.
\item Asymmetric: use separate context binnings for windows of various distance from current position - the further, the stronger reduction, e.g. \texttt{bin12near[}$x_{i-1},x_{i-2}$\texttt{]} into 64 bins and \texttt{bin12far[}$x_{i-3},x_{i-4}$\texttt{]} into 8 bins. To maintain the speed, the latter can be obtained by cutting bits from already calculated former e.g. \texttt{bin12far[]=bin12near[]>>3}, what can be optimized by bin enumeration. For example we first find large binning for near values using $\textrm{Pr}(x_i|x_{i-1},x_{i-2})$ distribution, then on these bins perform second binning  based on $\textrm{Pr}(x_i|x_{i-3},x_{i-4})$ evaluation, finally grouping e.g. 8 small into 1 large.
\item Hierarchical: build tree of contexts - binning of further values depending on current nearer values, e.g. \texttt{bin12far[]} depending also on outcome of \texttt{bin12near[]} - what seems more appropriate from statistics perspective, but has much higher cost. To construct it, after building tree for near, restrict windows $W$ to subsets corresponding to each bin, then build separate tree for each, and so on until e.g. some cost threshold.
\end{enumerate}

\subsection{Hidden state context models (HSCM)}
Discussed higher order approaches might require multiple table uses per step, bringing a natural question of squeezing comparable behavior into a single optimized table.

For example by evolving, based on value $\texttt{val}=x_i$, a hidden state acting as context - determining probability distribution for the next value $\textrm{Pr}(x_i|c)\approx \textrm{Pr}(x_i|\texttt{state})$:
\be \texttt{state = transition[state,val]} \label{HSCM} \ee

This function could e.g. contain some counter, e.g. counting numbers of changes of quality scores like in FQZComp \texttt{delta}~\cite{cram} - including long-range dependencies. To include some additional context like position, it can be put into \texttt{val} in (\ref{HSCM}), grouped or binned with the previous value. We can also use pair: (\texttt{state}, other context), e.g. grouped or binned, to determine probability distribution for the next value.

It is tempting to just use context binning to choose this transition function, however, it would require to find a fixed point of such operation (based on symbol sequence), what seems quite difficult. There is further mentioned machine learning-like approach for such optimization, let us first discuss inexpensive approach with state packing window of binned previous values.\\

\subsubsection{Hierarchical context binning (HCB)}
\begin{figure}[t!]
    \centering
        \includegraphics{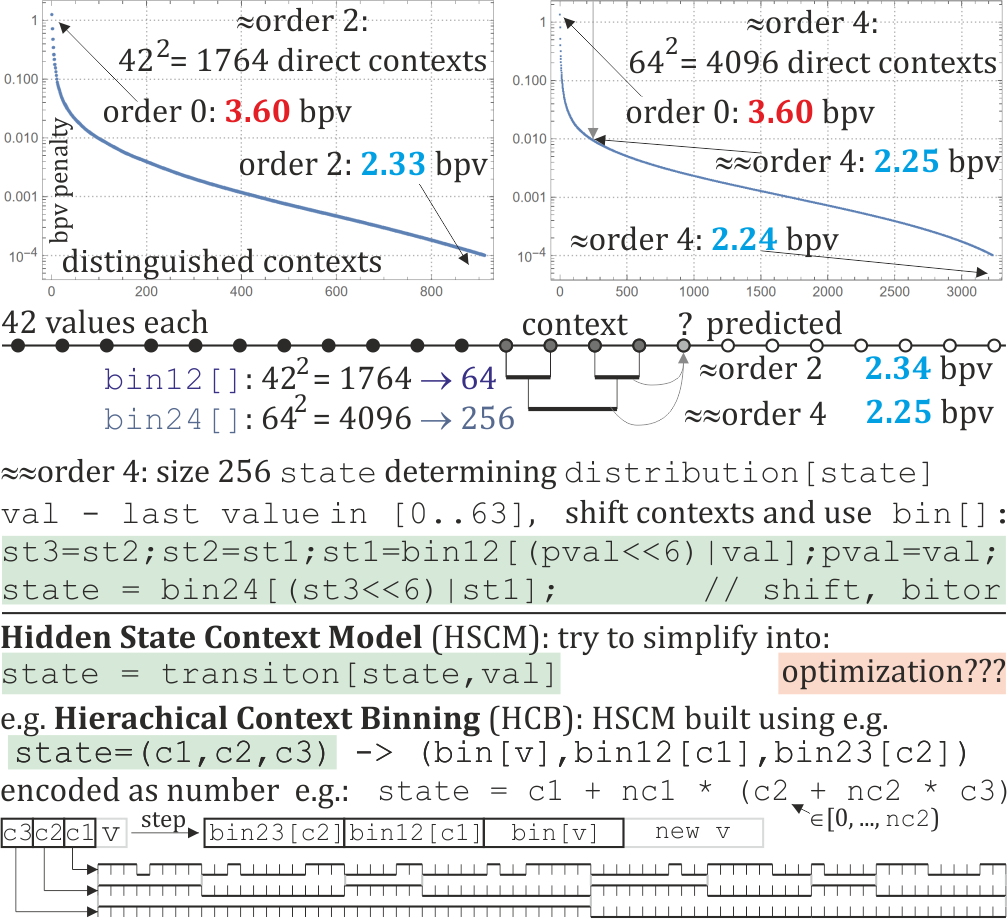}
        \caption{Nested context binning for approximation of high order models (for quality scores), e.g. order 2 with 1 table use per symbol, order 4 with 2 table uses, order 8 with 3 and so on. For example \texttt{bin12[]} here extracts crucial information from 2 neighboring values as one of 64 states, then \texttt{bin24[]} groups them further into approximated representation of order 4 context as one of 256 states - to be used to choose probability distribution for entropy coding of the current value, e.g. as modification of fast order 1 rANS. It is tempting to try to simplify this approach into single table use (HSCM), with HCB example of \texttt{state} encoding window of binned previous values.}
       \label{hier}
\end{figure}

The (\ref{HSCM}) transition can be inexpensively optimized for state packing moving window of binned a few previous values, if this binning was made in hierarchical way as in bottom of Fig. \ref{hier}: predicting $x_t$, the previous value is used as $\texttt{bin}[x_{t-1}]$, earlier as $\texttt{bin12}[\texttt{bin}[x_{t-2}]]$, then $\texttt{bin23}[\texttt{bin12}[\texttt{bin}[x_{t-3}]]]$, and so on - choosing further as binning from one position closer (can be trivial e.g. \texttt{bin23[x]=x}).

Choosing binning in such hierarchical way, we can easily shift this window into the next position, e.g.:
$$\texttt{(c1,c2,c3) -> (bin[v],bin12[c1],bin23[c2])}$$
We can encode such binned context window into a single natural number e.g. using numeral systems:
$$\texttt{state = c1 + nc1 * (c2 + nc2 * (c3 + ...))}$$
where $\texttt{c1}\in \{0,\ldots,\texttt{nc1}-1\}$, \texttt{nc} are numbers of bins for succeeding used positions. Then division by \texttt{nc} and taking modulo allows to uniquely decode \texttt{(c1, c2, c3, ...)}. For convenient power-of-2 number of states, all \texttt{nc} would need to be power-of-2.

So we choose binnings for succeeding positions using Algorithm \ref{bin}, starting with binning for the previous position (or just build one tree and cut it in different levels). Then find \texttt{transition} function: by considering all \texttt{(state,val)} pairs, decoding state, shifting to next position using \texttt{val}, and encoding back \texttt{state} - for update \texttt{state=transition[state,val]}.

This way we can approximate long range context models, especially if using (identical) binnnings into 2 bins from some position (e.g. high/low quality score) - with $2^{order}$ growth of number of states. This order can be further increased e.g. by grouping multiple symbols into one, like 3 nucleotides in Fig. \ref{binning}. We could also include some additional mechanisms into such state, e.g. extending it by a few bits working as counter, summator, or a more complex function of values in the \texttt{state}.\\

\subsubsection{Continuous $\rightarrow$ deterministic machine learning-like optimization} General HSCM could also allow to include long-range dependencies, e.g. counter like \texttt{delta} in FQZComp~\cite{cram}. There might be considered some semi-automatic search/optimization, e.g. trying to include such counters or more complex types of behavior.

For automatic optimization it seems useful to first transform it into continuous optimization: of probability distributions of states after \texttt{transition} $t$ (probabilistic instead of deterministic) - allowing to use gradient optimization methods like in machine learning approaches. \\

\noindent E.g. using softmax parametrisation $t,d$ of distributions:
$$(T_x)_{rs}:=\textrm{Pr}(s_i=s|s_{i-1}=r,x_{i-1}=x)=\exp(t_{srx})$$
$$\textrm{Pr}(x_i=x|s_i=s)= \exp((d_x)_s)$$
$$\textrm{satisfying:}\quad\forall_{rx}\sum_{s}\exp(t_{srx})=1\qquad \forall_s \sum_{x}\exp(d_{xs})=1 $$
$$P_{is}\equiv (P_i)_s:=\textrm{Pr}(s_i=s)\qquad\qquad P_{i+1}=P_i\cdot T_{x_i} $$
For optimized evaluation function:
\be \arg\max_{td} F\quad\textrm{for}\quad F=\sum_{si} P_{is}\,\ln(\textrm{Pr}(x_i|s_i))=\sum_i P_i\cdot d_{x_i} \label{optm}\ee
being minus expected number of nits, e.g. also to be summed over DNA reads. Starting states can be chosen fixed $P_1=(1,0,\ldots,0)$, initial distributions as uniform ($t,d\,=\,$const). The $d$ can be removed estimating it from $P_{is}$:
$$\textrm{Pr}(x|s)=\frac{\textrm{Pr}(s|x)\, \textrm{Pr}(x)}{\textrm{Pr}(s)}=\frac{\sum_{i:x_i=x} P_{is}}{\sum_i P_{is}}$$

To get deterministic transitions ($t_{srx}\in \{0,1\}$), we can try greedy approach: after optimizing (\ref{optm}), find maximal transition probability, then fix it to deterministic transition $t_{srx}=1$, $\forall_{s'\neq s} t_{s'rx}=0$. And so on: interleaving continuous optimization (\ref{optm}) with already fixed transitions, and fixing the highest probability transition, until all $t$ are deterministic.

To summarize, the choice of general HSCM is a promising direction with various optimization opportunities for both speed and compression ratio, which better understanding requires further research. Discussed model reduction should also help with overfitting, providing more universal, interchangeable models.

\section{Varying models}
The second discussed optimization direction is varying of models ($M$, of various types) - trying to better adapt to local statistics. Standard briefly mentioned here direction is adaptivity: using evolving models: $M_i$ for position $i$, with transition $M_i\to^{x_i} M_{i+1}$ based on processed symbol $x_i$.

As in Fig. \ref{qual} and \ref{bases}, here we will focus on proposed model clustering: optimizing some relatively small set of models $\{M_j\}_{j=1..k}$ (as centroids of clusters) and switching between them, e.g. for each "read" - writing in its header which of $k$ models to use (or maybe using some more complex switching mechanism).

\subsection{Model clustering}
Sequencing data usually contains large number of reads - of lengths from dozens to thousands of nucleotides (depending on technique). The reads are supposed to be finally aligned, what would e.g. allow to encode differences from such aligned consensus sequence - allowing for large compression ratio improvements (at least for bases).

Here let us assume we have just a set of reads as in standard FASTA/FASTQ file - treated independently. As there is usually freedom of read order, we could save lg((number of reads)!) bits e.g. by storing some information in their order. Each of these reads correspond to some local situation - might have different statistics.

The standard approach is using the same statistical model for each read, here it is proposed to group them into a few clusters of similar statistics and e.g. inform in each header which of these models to use. Encoder tests all and chooses the best one, decoder reads which should be used and decodes using it.

Algorithm \ref{kmeans} contains abstract pseudocode for standard k-means clustering algorithm~\cite{kmeans} adopted for this application: loop interlacing assignment of all reads to the closest centroid (best model here), and optimizing centroid (model here) for all members of its current cluster, until some convergence condition.

\begin{algorithm}[htbp]
\footnotesize{
\caption{k-means clustering of models}
\label{kmeans}
\begin{algorithmic}
\STATE Initialize: choose random $k$ reads, calculate $M_j$ model for $j$-th of them
\WHILE{convergence condition e.g. change below some threshold}
\FOR{each read}
\STATE calculate cost in bits for using each of $k$ models for this read
\STATE find minimal cost: $j$-th model, assign this read to $R_j \subset R$ subset
\ENDFOR
\FOR{each subset $R_j$}
\STATE calculate $M_j=$ model optimized for $R_j$ subset of reads
\ENDFOR
\ENDWHILE
\end{algorithmic}
}
\end{algorithm}
It interleaves evaluation of models (assigning to the best ones), and optimizing models for subsets of assigned e.g. reads (preferably by some merging) - as being quite general, it can be applied also to very complex models.

\begin{figure}[t!]
    \centering
        \includegraphics{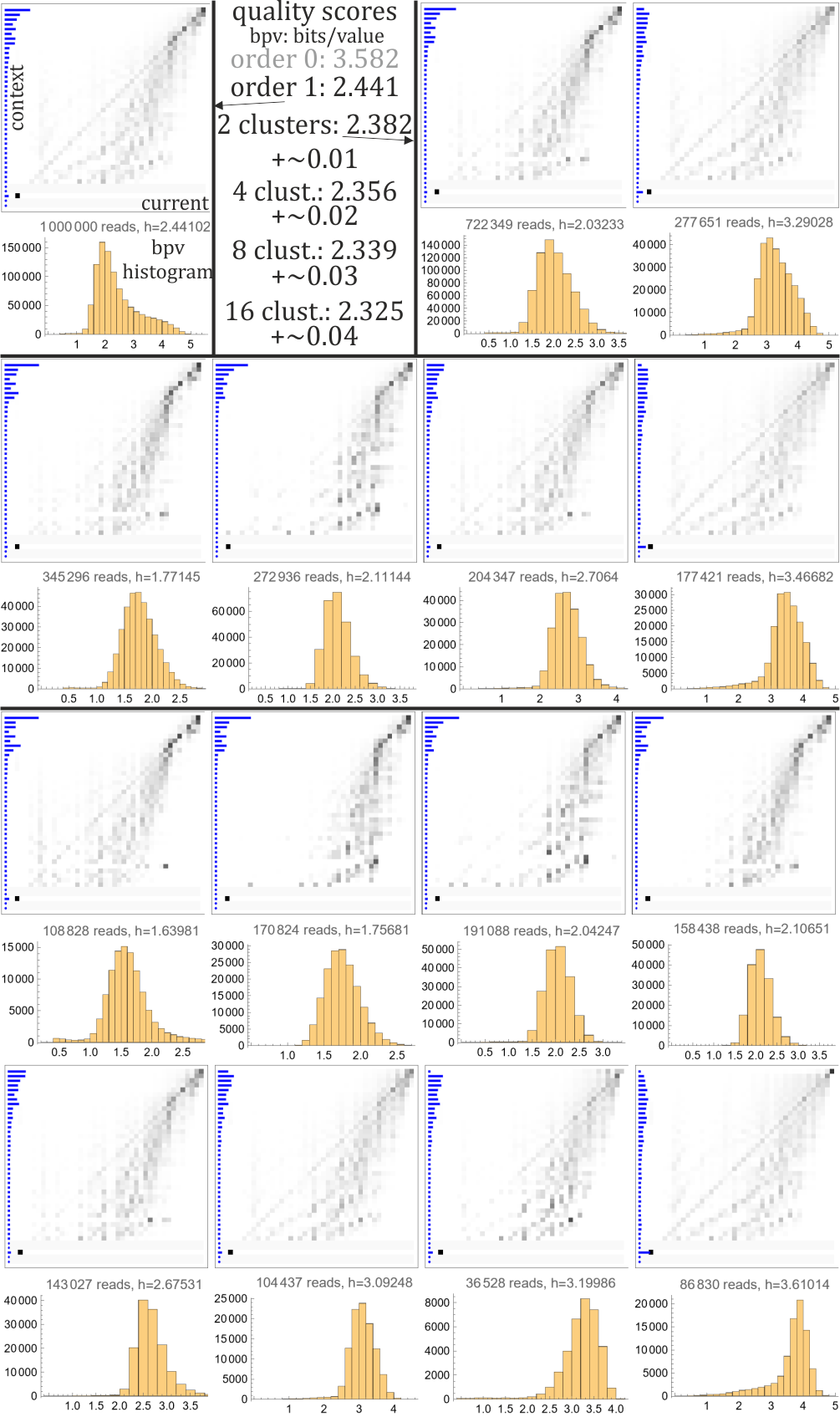}
        \caption{Model clustering of order 1 models for quality scores based on first 1 million reads of length 101 of ERR174310 - diagrams with context probabilities (blue bars), conditional distributions and bits/value histograms for 1,2,4,8 clusters found with k-means clustering, with numbers of reads for which given model leads to the smallest number of bits/value. We can observe split especially into low and high quality reads. We need to add which-model information in header of each read, which costs $\lg(k)$ if writing directly, or a bit more if using entropy coder, leading to $\approx 0.01-0.04$ additional bpv.}
       \label{qual}
\end{figure}

\begin{figure}[t!]
    \centering
        \includegraphics{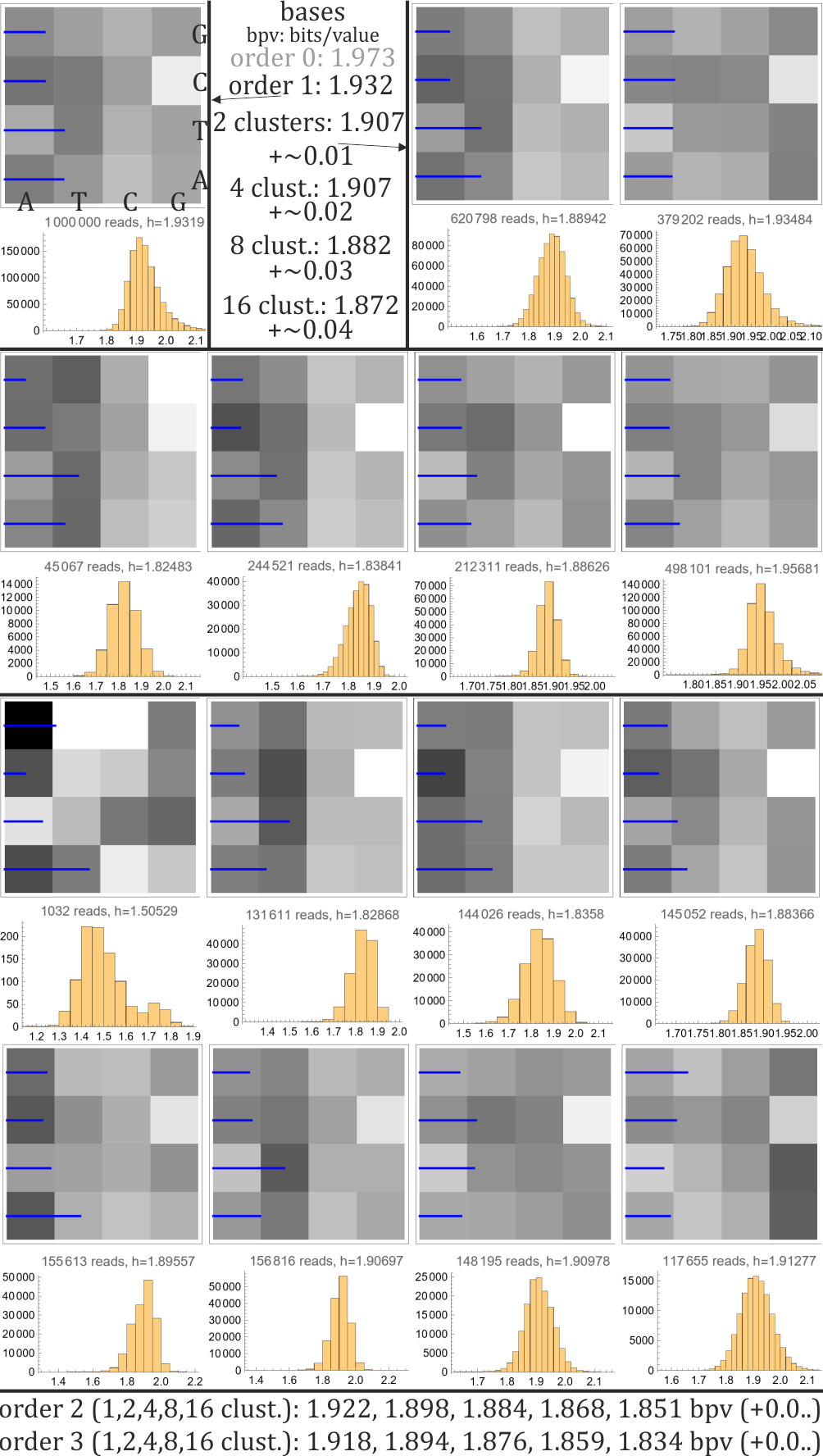}
        \caption{Model clustering  of order 1 models for nucleotides (bases) for first 1 million reads of ERR174310. Additional which-model information costs $\approx 0.01-0.04$ bits/value.}
       \label{bases}
\end{figure}

For example we can fix set of distinguished contexts $C$ e.g. as $l$ previous values in order $l$ model (order 1 in Fig. \ref{qual}, \ref{bases}), or its reduced size binned version from the previous section. Then for each read  calculate conditional (or joint) distributions ($p_c,P_c$), and for subsets of reads take their averages weighted with lengths.

CRAM FQZComp uses 2bit \texttt{select} in context as similar mechanism. Clustering e.g. k-means allows for its automatic optimization, also using 4 separate models should be faster as focusing on smaller tables than single 4x larger model.
\subsection{Adaptivity for non-stationary sources}

\begin{figure}[t!]
    \centering
        \includegraphics{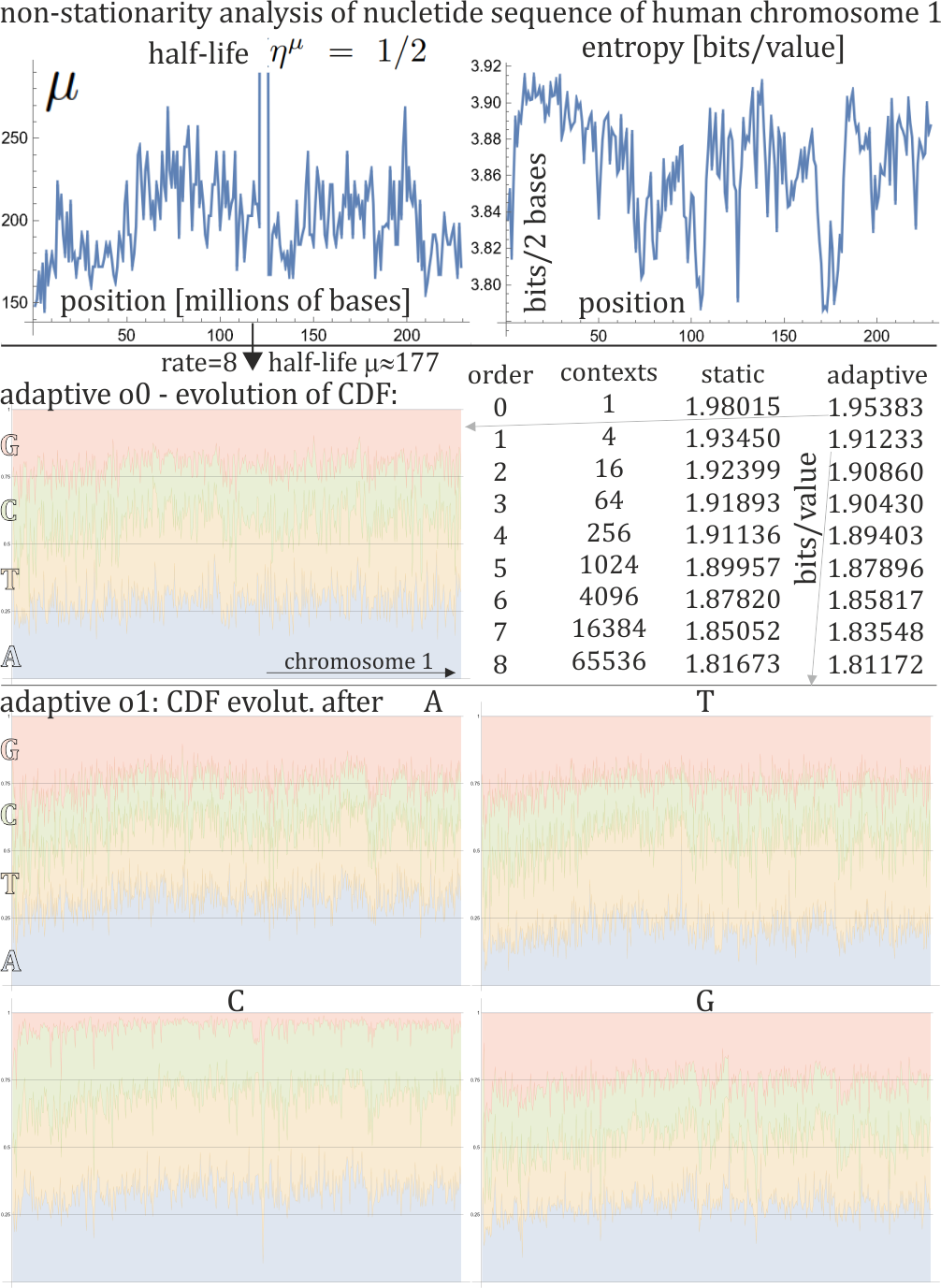}
        \caption{Top:Simple non-stationarity analysis of nucleotide sequence of human chromosome 1. Each block of 1 million bases was treated independently - there was performed search of $\eta$ forgetting rate of exponential moving average (\ref{ema}) leading to the best agreement: lowest bits/value entropy. There are presented its half-lifes $\mu=-1/\lg(\eta)$ (left), and this lowest entropy per pair of neighboring bases. Bottom: evolution of probabilities (as CDF) for adaptive order 0 and order 1 models - using (\ref{upd1}) for \texttt{rate}=8 ($\mu \approx 177$): applying and updating $\textrm{CDF}_c$ distribution for the current context $c$. There are also written tested bpv for stationary and adaptive models of order 0-8.}
       \label{nonst}
\end{figure}

More standard approach for varying models is adaptivity. There are very general and powerful machine learning approaches like LSTM~\cite{lstm}, but they are also extremely expensive - for practical data compression we should try to extract crucial behavior for inexpensive models. The basic approach is replacing averaging like $v=\textrm{mean}(f(x_i))$, with "evolving averaging" - exponential moving average (EMA):
\be v_{i+1}=\eta v_i +(1-\eta)f(x_i) \to (1-\eta)\sum_{d \geq 0} \eta^d f(x_{i-d})\label{ema}\ee
There is a difficult question of choosing this $0<\eta<1$ forgetting rate (usually $>0.9$), intuitively describing memory length of the system: $\propto \eta^d$ contribution of $d$ values before. We can translate it to half-life: contributions in distance $\mu=-1/\lg(\eta)$ weakens twice: $\eta^\mu=1/2$. Top of Fig. \ref{nonst} shows behavior of such optimized half-lifes along human chromosome 1: it is usually $\approx 200$, but also with regions of shorter or longer memory.

EMA can be used in various ways, e.g. for evolution of covariance matrix in online-PCA~\cite{opca} and adaptive linear regression~\cite{ada}, or just of center and scale parameter for general exponential power distribution $\rho(x)\sim \exp(-|x|^\kappa)$ family~\cite{aepd}.

For genetic data we rather focus on discrete probability distributions, requiring EMA for  "evolving frequency count". For this purpose it is convenient to work with cumulative distribution function (CDF): $\textrm{CDF}_i = \sum_{j<i} \textrm{Pr}(i)$. It for example allows for started in 2015 in LZNA compressor\footnote{https://fgiesen.wordpress.com/2015/12/21/rans-in-practice/} popular very fast SIMD implementations of adaptive rANS for up to size 16 alphabet:
\be \textrm{CDF}\ +=\ (\textrm{mixCDF}-\textrm{CDF}) >> \textrm{rate} \label{upd}\ee
where rate corresponds to $-\lg(1-\eta)$, the mixCDF is CDF of recently processed block of symbols - such update literally shifts old distribution toward the new distribution.

In practice such update of entire probability distribution is usually used every symbol - what allows to be close to local (past) distribution, but is costly - it is worth to consider also sparser updates e.g. every 100 symbols, also allowing to practically update more complex models.

While the above is adaptive order 0 model, we can easily expand it into context adaptive models, e.g. higher order like in bottom of Fig. \ref{nonst}. For example maintaining separate $\textrm{CDF}_c$ for each considered context $c\in C$ (e.g. $l$-previous, binned, hidden, etc.), for each processed symbol there is recognized context $c\in C$, used $\textrm{CDF}_c$ probability distribution, which is later updated:
\be \textrm{CDF}_c\ +=\ (\textrm{mixCDF}_c-\textrm{CDF}_c) >> \textrm{rate} \label{upd1}\ee

\section{Conclusions and further work}
There were proposed especially context binning and model clustering approaches - allowing for parallel inexpensive optimizations e.g. of context based models. Additional decoding cost is low (up to a few table uses per symbol). Encoding can also remain inexpensive, especially if restricting to some globally optimized models (e.g. for each model of DNA sequencer).

This is initial version of article to be expanded in the future (also applied in real compressors), for example:

\begin{itemize}
  \item Bases and quality scores were treated here independently, while in practice they are slightly correlated. To include their statistical dependencies for improving compression ratio, we can e.g. pack both into single value like \texttt{(qscore<<2)|base}.
  \item The reads were treated as independent, while in practice each base usually appears in multiple reads. To include their dependence we can e.g. first perform alignment to find consensus sequence (as looking most probable real sequence based on all the reads), then encode only differences from this consensus. Finally to encode this consensus, we can encode differences from the nearest reference sequence e.g. of humans.
  \item Mentioned Hidden State Context Model (HSCM) seems a promising development direction - allowing to include also far contexts in very inexpensive way (including advantages of RLE). However, it needs research for practical optimizations, approximations, maybe semi-manual.
  \item Discussed model clustering switches between a few models - it might be worth to develop mechanisms automatically switching between them based also on processed symbols, e.g. some adaptation with attraction to centroids.
\end{itemize}

\textbf{Finally}, we could prepare default e.g. 20 statistical models optimized based on a large dataset for a given model of sequencer. Then compressor tests on given file and chooses e.g. 4 best models - marked in file header. Then for each read writes 2 bits choosing one of these 4 models, allowing inexpensive parallel processing of reads.

Such models can be chosen by first identifying large potentially valuable context, like previous values, position in read, and then trying to squeeze it into inexpensive approximation e.g. using discussed context binning techniques.

\bibliographystyle{IEEEtran}
\bibliography{cites}

\begin{thebibliography}{10}
\providecommand{\url}[1]{#1}
\csname url@samestyle\endcsname
\providecommand{\newblock}{\relax}
\providecommand{\bibinfo}[2]{#2}
\providecommand{\BIBentrySTDinterwordspacing}{\spaceskip=0pt\relax}
\providecommand{\BIBentryALTinterwordstretchfactor}{4}
\providecommand{\BIBentryALTinterwordspacing}{\spaceskip=\fontdimen2\font plus
\BIBentryALTinterwordstretchfactor\fontdimen3\font minus
  \fontdimen4\font\relax}
\providecommand{\BIBforeignlanguage}[2]{{%
\expandafter\ifx\csname l@#1\endcsname\relax
\typeout{** WARNING: IEEEtran.bst: No hyphenation pattern has been}%
\typeout{** loaded for the language `#1'. Using the pattern for}%
\typeout{** the default language instead.}%
\else
\language=\csname l@#1\endcsname
\fi
#2}}
\providecommand{\BIBdecl}{\relax}
\BIBdecl

\bibitem{prec}
E.~A. Ashley, ``Towards precision medicine,'' \emph{Nature Reviews Genetics},
  vol.~17, no.~9, pp. 507--522, 2016.

\bibitem{BW}
M.~Burrows and D.~Wheeler, ``A block-sorting lossless data compression
  algorithm,'' in \emph{Digital SRC Research Report}.\hskip 1em plus 0.5em
  minus 0.4em\relax Citeseer, 1994.

\bibitem{cram}
\BIBentryALTinterwordspacing
J.~K. Bonfield, ``Cram 3.1: Advances in the cram file format,''
  \emph{Bioinformatics}, 2022. [Online]. Available: \url{CRAM 3.1: Advances in
  the CRAM File Format}
\BIBentrySTDinterwordspacing

\bibitem{AC}
J.~Rissanen and G.~G. Langdon, ``Arithmetic coding,'' \emph{IBM Journal of
  research and development}, vol.~23, no.~2, pp. 149--162, 1979.

\bibitem{ANS}
J.~Duda, K.~Tahboub, N.~J. Gadgil, and E.~J. Delp, ``The use of asymmetric
  numeral systems as an accurate replacement for huffman coding,'' in
  \emph{2015 Picture Coding Symposium (PCS)}.\hskip 1em plus 0.5em minus
  0.4em\relax IEEE, 2015, pp. 65--69.

\bibitem{kmeans}
J.~MacQueen \emph{et~al.}, ``Some methods for classification and analysis of
  multivariate observations,'' in \emph{Proceedings of the fifth Berkeley
  symposium on mathematical statistics and probability}, vol.~1, no.~14.\hskip
  1em plus 0.5em minus 0.4em\relax Oakland, CA, USA, 1967, pp. 281--297.

\bibitem{lstm}
F.~A. Gers, N.~N. Schraudolph, and J.~Schmidhuber, ``Learning precise timing
  with lstm recurrent networks,'' \emph{Journal of machine learning research},
  vol.~3, no. Aug, pp. 115--143, 2002.

\bibitem{opca}
M.~K. Warmuth and D.~Kuzmin, ``Randomized online pca algorithms with regret
  bounds that are logarithmic in the dimension,'' \emph{Journal of Machine
  Learning Research}, vol.~9, no. Oct, pp. 2287--2320, 2008.

\bibitem{ada}
J.~Duda, ``Parametric context adaptive laplace distribution for multimedia
  compression,'' \emph{arXiv preprint arXiv:1906.03238}, 2019.

\bibitem{aepd}
------, ``Adaptive exponential power distribution with moving estimator for
  nonstationary time series,'' \emph{arXiv preprint arXiv:2003.02149}, 2020.

\end{thebibliography}
\end{document}